\documentclass[10pt,preprint,twocolumn]{aastex7}
\newcommand{\roma}[1]{\uppercase\expandafter{\romannumeral#1}}
\newcommand{\speed}[1]{#1 km~s${}^{-1}$}

\newcommand{\nfig}[1]{Figure~\ref{#1}}

\usepackage{amsmath}
\usepackage{subfigure}
\usepackage{hyperref}

\usepackage{lineno}
\usepackage{color}
\usepackage{makecell}
\usepackage{graphicx}
\usepackage{natbib}
\usepackage{amssymb,txfonts}
\usepackage{multirow}
\usepackage{array}
\usepackage{sidecap}
\usepackage{hyperref}
\usepackage{epstopdf}
\usepackage{footnote}
\usepackage{tabularx}
\usepackage{booktabs}

\shorttitle{Formation of Counterstreaming Mass Flows and Doppler Bullseye Patterns}
\shortauthors{Zhou et al.}
\received{January 31, 2025}
\revised{April 17, 2025}
\accepted{\today}

\begin{document}
\title{A New Formation Mechanism of Counterstreaming Mass Flows in Filaments and the Doppler Bullseye Pattern in Prominences}
\correspondingauthor{Yuandeng Shen}

\author{Chengrui Zhou$^{1,6}$}
\noaffiliation{}
\affiliation{Yunnan Observatories, Chinese Academy of Sciences, Kunming 650216, China}
\affiliation{School of Aerospace, Harbin Institute of Technology, Shenzhen 518055, China}
\affiliation{Shenzhen Key Laboratory of Numerical Prediction for Space Storm, Harbin Institute of Technology, Shenzhen 518055, China}
\affiliation{State Key Laboratory of Solar Activity and Space Weather, National Space Science Center, Chinese Academy of Sciences, Beijing 100190, China}
\affiliation{School of Physics and Astronomy, Yunnan University, Kunming 650500, China}
\affiliation{University of Chinese Academy of Sciences, Beijing, 100049, China}
\email{zhouchengrui@ynao.ac.cn}  

\author[orcid=0000-0001-9493-4418]{Yuandeng Shen}
\affiliation{School of Aerospace, Harbin Institute of Technology, Shenzhen 518055, China}
\affiliation{Shenzhen Key Laboratory of Numerical Prediction for Space Storm, Harbin Institute of Technology, Shenzhen 518055, China}
\affiliation{State Key Laboratory of Solar Activity and Space Weather, National Space Science Center, Chinese Academy of Sciences, Beijing 100190, China}
\email[show]{ydshen@hit.edu.cn}

\author{Chun Xia}
\affiliation{School of Physics and Astronomy, Yunnan University, Kunming 650500, China}
\email{chun.xia@ynu.edu.cn}

\author{Dongxu Liu}
\affiliation{Yunnan Observatories, Chinese Academy of Sciences, Kunming 650216, China}
\affiliation{University of Chinese Academy of Sciences, Beijing, 100049, China}
\email{liudongxu@ynao.ac.cn}

\author{Zehao Tang}
\affiliation{Yunnan Observatories, Chinese Academy of Sciences, Kunming 650216, China}
\affiliation{University of Chinese Academy of Sciences, Beijing, 100049, China}
\email{tangzh@ynao.ac.cn}

\author{Surui Yao}
\affiliation{Yunnan Observatories, Chinese Academy of Sciences, Kunming 650216, China}
\affiliation{University of Chinese Academy of Sciences, Beijing, 100049, China}
\email{yaosurui@ynao.ac.cn}

\begin{abstract}
The eruption of solar prominences can eject substantial mass and magnetic field into interplanetary space and cause geomagnetic storms. However, various questions about prominences and their eruption mechanism remain unclear. In particular, what causes the intriguing Doppler bullseye pattern in prominences has not yet been solved, despite some preliminary studies proposing that they are probably associated with counterstreaming mass flows. Previous studies are mainly based on single-angle and short timescale observations, making it difficult to determine the physical origin of Doppler bullseye patterns in prominences. Here, {taking advantage} of stereoscopic observations taken by the {\em Solar Dynamics Observatory} and the {\em Solar Terrestrial Relations Observatory} and a three-dimensional numerical simulation, we investigate the origin of prominence Doppler bullseye pattern by tracing a long-lived transequatorial filament/prominence from July 23 to August 4, 2012. We find that repeated coronal jets at one end of the prominence can launch the Doppler bullseye pattern. It is evidenced in our observations and simulation that during the forward traveling of jet plasma along the helical magnetic field structure of the prominence, part of the ejecting plasma can not pass through the apex of the prominence due to the insufficient kinetic energy and therefore forms a backward-moving mass flow along the same or neighboring magnetic field lines. This process finally forms counterstreaming mass flows in on-disk filaments. When the on-disk filament rotates to the solar limb to be a prominence, the counterstreaming mass flows are naturally observed as a Doppler bullseye pattern.
\end{abstract}

\section{Introduction}\label{intro}
A solar prominence is often a large and cold mass-containing magnetic field structure in the hot corona with its two ends rooted in the photosphere; it appears as a bright emission feature at the solar disk limb but dark elongated absorption structures on the disk; astronomers refer to the two different observational features respectively as prominences and filaments, although they are the same entity in the solar atmosphere \citep{1998SoPh..182..107M, 2007ApJ...667L.105J, Mackay2010, 2015ApJ...814L..17S, 2018ApJ...863..192L, 2024ApJ...970..110G}. Previous studies show that the basic {magnetic structure of filaments} could be sheared arcade~\citep{kip57,lite05, Mackay2010, Aulanier1998a} or twisted magnetic flux ropes (MFRs)~\citep{2022ApJ...934L...9L, rust96, zhangj12, Zhou2017,2024ApJ...964..125S}, and their eruptions are likely to cause geomagnetic storms and catastrophic space weather in the near-Earth space environment {\citep[e.g.,][]{Mackay2010, 2011RAA....11..594S, 2012ApJ...750...12S,2021ApJ...923...45Z}}. It has widely been accepted that the magnetic field plays a key role in solar filaments' formation, stability, and eruption. However, to date, knowledge of the fine magnetic structure of the filaments is still scarce, and many questions about filaments remain unsolved.

High spatial resolution observations reveal that filaments consist of numerous thin thread-like structures \citep{2005SoPh..226..239L, 2015ApJ...814L..17S, 2018ApJ...863..192L, 2024ApJ...970..110G}. Since plasma is frozen on magnetic field lines in low $\beta$ plasma environments, these thin filament threads exhibit the mass and magnetic field distribution in filaments. The magnetic field lines that carry filament mass typically have a {concave upward} shape that is often referred to as a magnetic dip, within which the filament thread remains in a quasi-static state. However, the mass in the magnetic dip will flow predominantly along the concave filament thread and sometimes forms longitudinal mass oscillation around the bottom of the dip~\citep{2014ApJ...795..130S, 2022MNRAS.516L..12T,2021A&A...647A.112Z}. If independent antiphase longitudinal oscillations exist simultaneously in neighboring filament threads, they might be observed as counterstreaming mass flows in filaments ~\citep{2003SoPh..216..109L, 2015ApJ...814L..17S, 2020NatAs...4..994Z}. 

\cite{1998Natur.396..440Z} reported counterstreaming mass flows in filaments along the spine and barbs. In recent years, high spatiotemporal resolution observations have evidenced the ubiquitous existence of this phenomenon in filaments~\citep[e.g.,][]{2015ApJ...814L..17S, 2018A&A...611A..64D, 2020ApJ...897L...2P} and active region magnetic loops~\citep{2019ApJ...881L..25Y}, but the physical origin of counterstreamings is still unclear. So far, solar physicists have proposed several candidate physical mechanisms to interpret the formation of counterstreaming mass flows. For example, siphonic effect owning to the pressure imbalance around the two footpoints of a filament~\citep{2014ApJ...784...50C}, small-scale reconnection events such as network jets~\citep{2020ApJ...897L...2P} and EUV brightenings~\citep{2019ApJ...881L..25Y}, opposite oscillation mass along different filament threads~\citep{2003SoPh..216..109L,2020NatAs...4..994Z}, and the generation of upward mass flows in vertical prominence threads which carry continuous downward mass flows due to the collapse of prominence bubbles~\citep{2015ApJ...814L..17S}. These studies demonstrated that the origin of counterstreaming mass flows in filaments might have different physical mechanisms. For a specific event, the origin of counterstreaming mass flows can be due to one or a combination of possible mechanisms. The mechanism of random heating at the footpoints of filaments due to magnetic reconnection has caught more attention from theorists, and it has been tested in several numerical simulation works to be a valuable way to drive counterstreaming mass flows in filaments~\citep{1991ApJ...378..372A, 2012ApJ...746...30L, 2020NatAs...4..994Z}. However, other possible mechanisms still need more theoretical support in addition to observations.
                
Over the solar disk limb, filaments are observed as prominences. Spectroscopic observations reveal the existence of the so-called bullseye pattern in prominence-containing cavities, which appear as concentric circles of alternate pattern of red- and blue-shift mass flows along the line-of-sight (LOS)~\citep{2013ApJ...770L..28B}. A bullseye pattern contains hot (million K) and persistent LOS velocity in the range of \speed{5-10}~\citep{1999ApJ...513L..83H,2013ApJ...770L..28B}. Restricted by the limit of limb observation, prominence bullseye patterns are interpreted as evidence of magnetic flux ropes (MFR) characterised by coherently twisted magnetic field lines winding around the same axis \citep{2020RAA....20..165L, 2024ApJ...975L...5Y}. The physical linkage between counterstreaming mass flows and prominence magnetic structure has not yet been established, despite this knowledge being critical to investigating the origin of bullseye patterns in prominences. 

In this paper, based on multi-angle and multi-wavelength observations, we analyze the origin of counterstreaming mass flows and the formation of the bullseye pattern in a long-lived filament/prominence. We found that continuous recurrent coronal jets around one footpoint of the filament can naturally launch counterstreaming mass flows in filaments and bullseye patterns in prominences. The observations and instruments are introduced in Section ~\ref{sec:data}. The jets' dynamic evolution and filament formation are presented in Section ~\ref{sec:result}. The numerical simulation setting and results are given in Section ~\ref{sec:numerical}. Conclusion and discussions are summarized in Section ~\ref{sec:summary}.

\section{Observations and Methods}\label{sec:data}
The filament under study was detected by various instruments, including the {\em Solar Dynamic Observatory} ~\citep[{\em SDO};][]{2012SoPh..275....3P}, the {\em Solar Terrestrial Relations Observatory}~\citep[{\em STEREO};][]{2004SPIE.5171..111W}, the Solar Magnetic Activity Research Telescope \citep[SMART;][]{2004ASPC..325..319U} and the Coronal Multi-Channel Polarimeter \citep[CoMP;][]{2008SoPh..247..411T}. The {\em SDO} observes full-disk solar images in 10 visible, ultraviolet (UV), and extreme-ultraviolet (EUV) wavelengths. The Atmospheric Imaging Assembly {\citep[AIA;][]{2012SoPh..275...17L}} onboard the {\em SDO} is equipped with various filters that detect plasma over a wide temperature range. We mainly use the AIA channels at 304 \AA\ (0.05 MK), 171 \AA\ (0.6 MK), 193 \AA\ (1.6 MK), 211 \AA\ (2.0 MK), and 1600 \AA(10000 K) in this study. The AIA full-disk multi-wavelength images have a pixel size of 0\arcsec.6, a cadence of 12 (24) s for EUV (UV) images. The full-disk LOS photospheric magnetograms and continuum intensity images taken by the Helioseismic and Magnetic Imager~{\citep[HMI;][]{2012SoPh..275..229S}}. The HMI has a pixel size of 0\arcsec.5~and a cadence of 45 s. The Solar Terrestrial Relations Observatory~{\citep[STEREO;][]{2008SSRv..136....5K} provides 195 \AA\ and 304 \AA\ full-disk images from the other observing angle. The {\em STEREO} 195 (304) \AA\ images have a cadence of 5 (10) minutes and a pixel size of 1\arcsec.6. We also used the full-disk H$\alpha$ observations taken by the SMART, which have a cadence of 2 minutes and a pixel size of 0\arcsec.6. The CoMP at the Mauna Loa Solar Observatory provides coronal velocity images that show critical information about the LOS Doppler velocity from the forbidden Fe XIII lines at  10747 \AA. The above instruments together provide us with a perfect observing condition to diagnose the formation of the observed counterstreaming mass flows and the Doppler bullseye pattern in the filament/prominence, as well as a series of repeated coronal jets around the one footpoint of the filament/prominence.

\begin{figure*}
\centering
\includegraphics[width=0.9\textwidth]{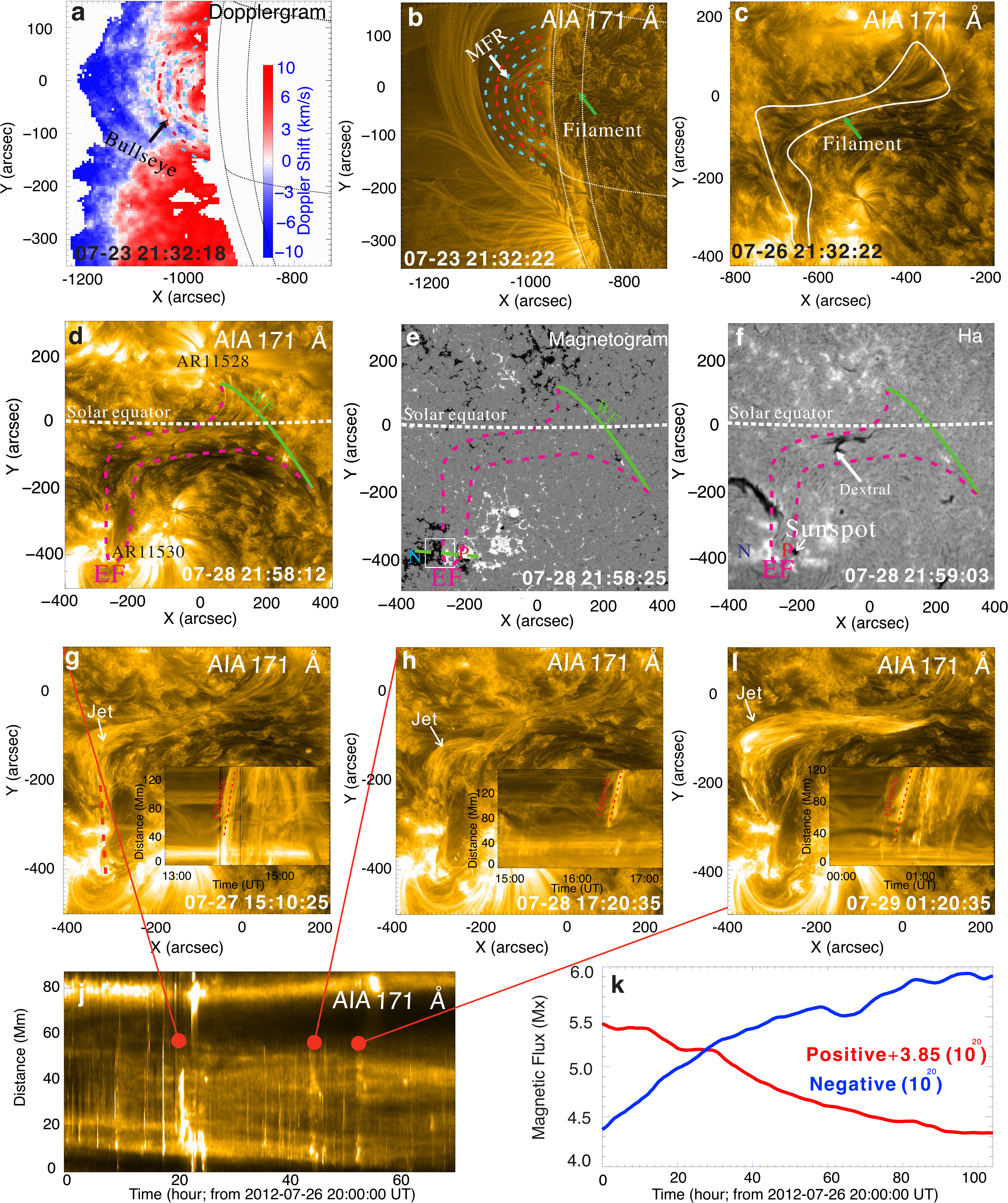}
\caption{The Doppler bullseye pattern, filament configuration, associated coronal jets, and magnetic flux variations. Panel (a) is a CoMP Doppler velocity map on July 23 showing the Doppler bullseye pattern. Panel (b) is an AIA 171 \AA\ image at the same time as the Doppler velocity map in panel (a), and the overlaid dashed curves are the same as those in panel (a). Panel (c) displays the filament on the disk on July 26, and the white curve outlines the location of the filament. Panels (d) -- (f) show the simultaneous AIA 171 \AA, HMI magnetogram, and H$\alpha$ images at about 21:58 UT on July 28. In each panel, the red dashed curves indicate the location of the filament, the green curve indicates the filament's west footpoint, and the white dotted curve indicates the position of the solar equator. Panels (g) -- (i) show three typical coronal jets close to the east footpoint of the filament, and the inset in each panel is the time-distance diagram made along the vertical red dashed line in panel (g). Panel (j) is a time-distance diagram made along the red dashed line in panel (g), which shows a large number of repeated coronal jets close to the east footpoint of the filament from 20:00 UT on July 26 to 20:00 UT on July 29. The three red lines with a red dot indicate the occurrence times of the three jets displayed in panels (g) -- (i). Panel (k) shows the variations of the positive (red) and negative (blue) magnetic fluxes within the source region of the coronal jets from 20:00 UT on July 26 to 20:00 UT on July 31.  (An animation of this event is available)}
\label{fig1}
\end{figure*}

\section{Observational Results}\label{sec:result}
We observed a long-lived intermediate filament across the solar equator from July 23 to August 4, 2012; it showed apparent counterstreaming mass flows and the Doppler bullseye pattern when observed on the solar disk and at the disk limb, respectively. The top row in \nfig{fig1} shows the filament at the east limb of the solar disk using the CoMP and 171 \AA\ images. \nfig{fig1} (a) shows a Dopplergram captured by the CoMP at 21:32:18 UT on July 23; it reveals some {alternate pattern of} rings of red- and blue-shift Doppler velocities (see the red and blue curves in \nfig{fig1} (a)). The blue- and red-shift Doppler velocity rings are indicative of plasma flows moving towards and away from the observer, showing a half-bullseye pattern in the field of view of the CoMP.  \nfig{fig1} (b) shows the filament in the AIA 171 \AA\ image at 21:32 UT on July 23, in which the red and blue guide lines agree with the pattern of the Doppler-shift signals in \nfig{fig1} (a). It reveals that the Doppler bullseye pattern observed in the CoMP Doppler images is located on the filament magnetic flux rope. \nfig{fig1} (c) shows the filament on the solar disk on July 26, 3 days after the limb observation in \nfig{fig1} (a) and (b). One can find that the filament was across the solar equator and connected active regions AR11530 and AR11528 in the south and north hemispheres, respectively. The second row in \nfig{fig1} shows the filament in the AIA 171 \AA, HMI LOS magnetogram, and the SMART H$\alpha$ images on July 28. In these images, the pink curves indicate the filament's position and shape. The eastern footpoint of the filament is compact and rooted in a positive magnetic region in AR11530. In contrast, the west footpoint is dispersed and distributed along an arc-shaped path in a quiet-Sun region with negative magnetic elements close to AR11528 (see the green curve in \nfig{fig1} (d) -- (f)). According to the orientation of the filament barb (see \nfig{fig1} (g)), it can be determined that the filament was a dextral one with negative helicity \citep{1998SoPh..182..107M, 2020RAA....20..166C}. 

We noticed that many coronal jets repeatedly occurred around the eastern footpoint of the filament, which can be easily observed in the AIA 171~\AA~images. We select three typical jets displayed in \nfig{fig1} (g) to (i), which show that the ejecting plasma of the jets was ejecting along the filament spine from the east to the west. Their speeds were \speed{129},  \speed{132} and \speed{101}, respectively (see the time-distance diagrams {in the inset}). To analyze these coronal jets' temporal and spatial relationship, we make a time-distance diagram using the AIA 171~\AA~images along the red dashed lines as shown in \nfig{fig1} (g). In the time-distance diagram, each vertical bright trajectory represents the forward ejecting hot plasma flow caused by a coronal jet. It can be seen in \nfig{fig1} (j) that there were plenty of coronal jets repeatedly occurring around the east footpoint of the filament and driving the westward-moving plasma flow along the filament spine. To investigate what caused the repeated occurrence of the coronal jets, we checked the variations of the positive and negative magnetic fluxes within the region around the source region of the jets (see the white rectangle in \nfig{fig1} (e)), and the result is plotted in \nfig{fig1} (k). It can be observed that the positive polarity flux experienced a continuous decline (red line), while the negative polarity flux underwent a gradual increase (blue line). Such a changing pattern of magnetic fluxes suggests that numerous repeated coronal jets were associated with the continuous flux cancellation between the positive and negative magnetic polarities~\citep{2007A&A...469..331J, shen12a, 2017ApJ...851...67S, 2017ApJ...844..131P, 2018ApJ...864...68S,shen21}.

\begin{figure*} 
\centering
\includegraphics[width=0.9\textwidth]{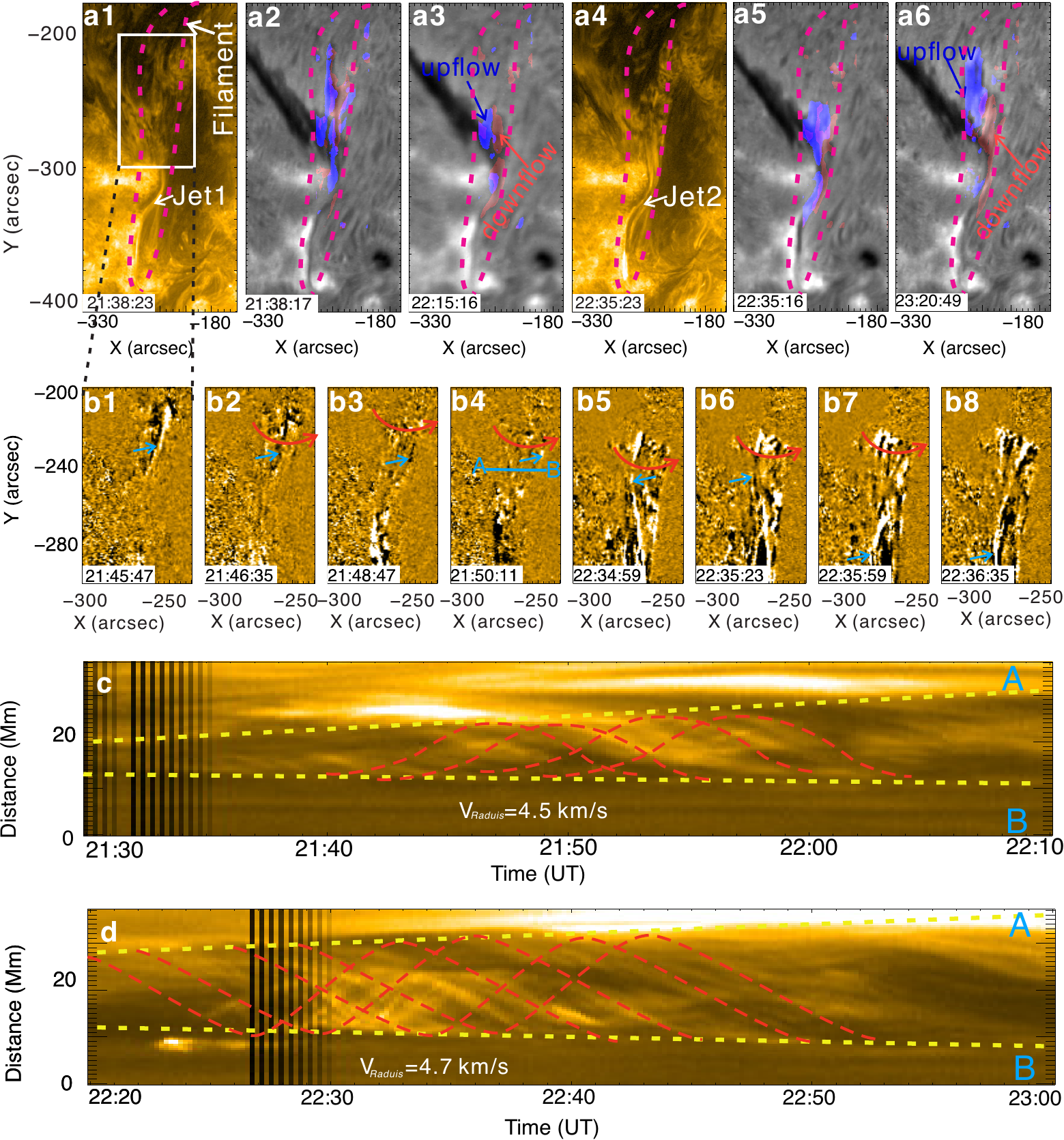}
\caption{The comprehensive analysis of the kinematic evolution of the filament internal flows driven by Jet1 and Jet2, and the associated movement patterns of filament threads. Panels (a1)--(a6) show the filament captured by AIA 171 \AA~images and composite tri-color H$\alpha$ images, where the blue color is the H$\alpha$-0.5 \AA~image, illustrating the filament upward mass flows, and the red color denotes the H$\alpha$+0.5 \AA~image, showing the filament downward mass flows. The red dashed curves in each panel delineate the scope of the filament. The white box in panel (a1) marks the field of view corresponding to panels (b1)--(b8). Panels (b1)--(b8): AIA 171 \AA~running difference images, where the blue arrows track the trajectory of the filament threads, and the red curved arrows indicate the rotational direction of the filament threads. Panels (c) and (d) present time-distance diagrams derived from the horizontal blue line in panel (b4). The yellow dashed lines highlight the inflation of the magnetic flux rope associated with Jet1 and Jet2. The expansion rates for Jet1 and Jet2 are \speed{4.5} and \speed{4.7}, respectively, and the red sinusoidal curves correspond to the trajectories of the rotational motions of the filament threads.}
\label{fig2}
\end{figure*}

We selected three conspicuous on-disk jets for detailed analysis; they occurred at about 21:30 UT and 22:26 UT on July 28 and 00:28 UT on July 29, respectively. For the sake of description, hereafter, we name the three jets as Jet1, Jet2, and Jet3, respectively. In principle, upward and downward mass flows in on-disk filaments can easily be detected by the off-band of the H$\alpha$ line \citep{2003PASJ...55..503M, 2014ApJ...786..151S}. To study the detailed kinematics of the mass flows driven by the jets in the filament, we made composite tri-color images by using the H$\alpha$-0.5 \AA\ (blue), H$\alpha$ line-center (grayscale image), and the H$\alpha$+0.5 \AA\ images (red). The results are displayed in \nfig{fig2} (a2), (a3), (a5) and (a6). In such composite tri-color images, the red and blue features represent the filament's upward and downward mass flows, respectively. Jet1 started at about 21:30 UT, and was well developed at 21:38 UT(see \nfig{fig2} (a1)), which excited an internal upward flow in the filament and caused a blue-shift feature in the H$\alpha$-0.5 \AA\ images  (see \nfig{fig2} (a2)). A few tens of minutes later, we observed a red-shift downward flow along the filament in the H$\alpha$ +0.5 \AA\ images, which might be caused by the falling back of some ejecting plasma (see \nfig{fig2} (a2)). Jet2 started at about 22:26 UT at the same location as Jet1, and its upward flows can be clearly identified at 22:35 UT (see \nfig{fig1} (a4)). Similar upward blue-shift and downward red-shift features also appeared in the H$\alpha$-0.5 \AA\ and +0.5 \AA\ images at different times (see \nfig{fig2} (a5) and (a6)). The composite H$\alpha$ images suggest that the upward and downward flows in the filament driven by Jet1 and Jet2 did not overlap; instead, the downward flows occurred to the right of the upward ones. As the filament plasma is frozen on the filament magnetic field lines, the separation of the upward and downward mass flows may reflect the helical magnetic structure of the filament's magnetic field. \nfig{fig2} (b1) -- (b8) show the AIA 171~\AA\ running difference images that show the moving features in the time sequence images. Here, a running difference image is created by subtracting the image at the previous moment from the present one. As the traveling of the hot jet plasma in the filament, the filament threads exhibited an obvious transverse motion in the AIA 171 \AA\ time sequence running difference images (see the blue arrows in \nfig{fig2} (b1) -- (b8)). In addition, the filament also showed a rotation motion at the same time (see the red arrows in \nfig{fig2} (b2) -- (b7)). We create two time-distance diagrams along the path across the filament (see the blue line in \nfig{fig2} (b4)) to show the transverse dynamic kinematics of the filament resulting from the two jets. Based on the time-distance plots shown in \nfig{fig2} (c) and (d), one can see that during the passage of jet1 and jet2, the filament expanded gradually in the lateral direction at a speed of about \speed{4.5 and 4.7}, respectively. In addition, some of the rotating or unwinding filament threads can also be observed (see the red dotted curves in \nfig{fig2} (c) and (d)). Such an intriguing kinematic is similar to unwinding coronal jets caused by the transfer of magnetic twists from close magnetic flux ropes to open magnetic field lines through magnetic reconnection \citep{2011ApJ...735L..43S}, and the sequential expansion of the threads might excite quasi-periodic fast-propagating magnetosonic wave trains \citep{2019ApJ...873...22S, 2022SoPh..297...20S, 2024ApJ...974L...3Z}.

\begin{figure*} 
\centering
\includegraphics[width=0.9\textwidth]{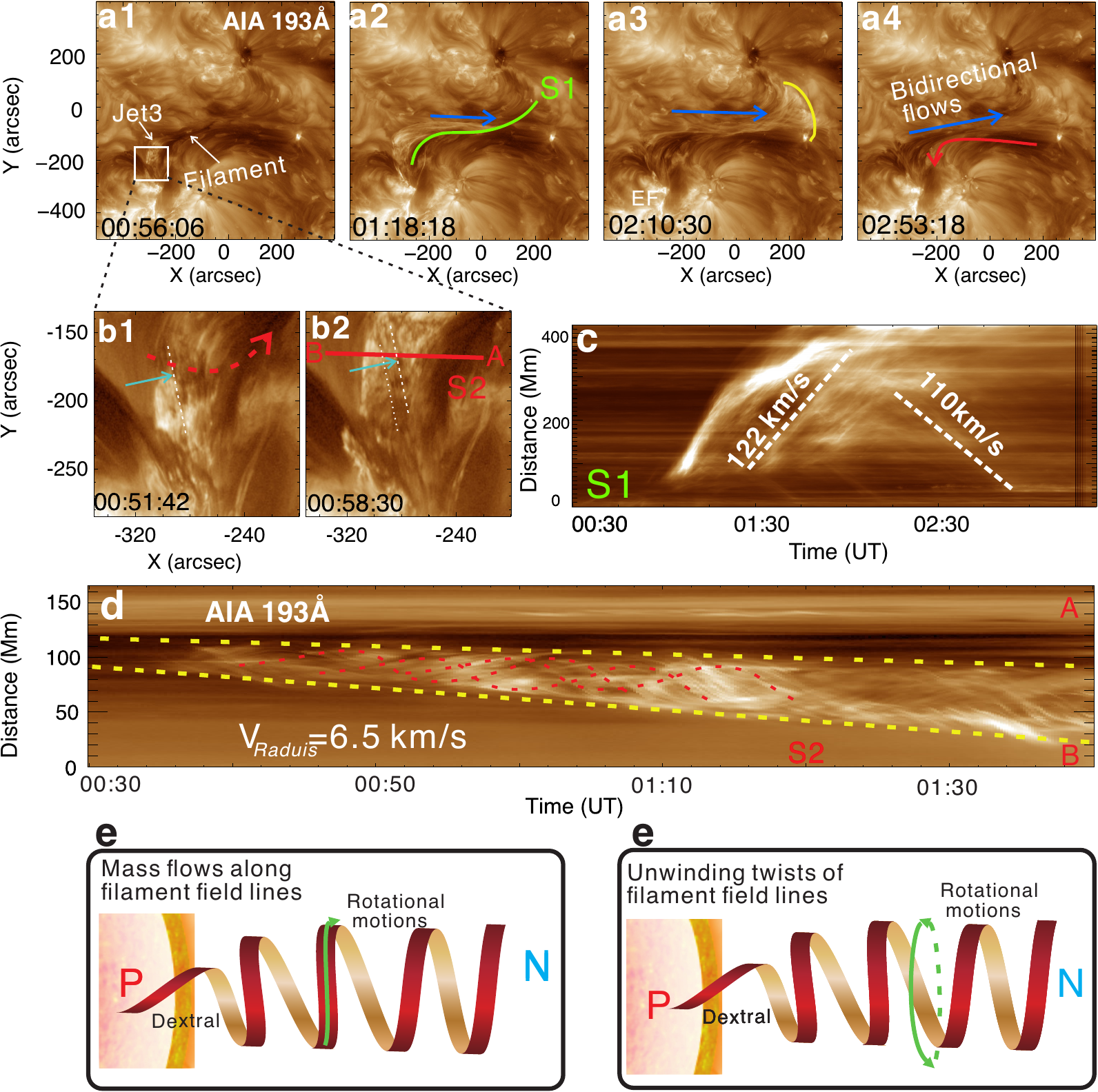}
\caption{The detailed kinematic analysis of the filament internal flows driven by Jet3 and the associated rotational pattern of filament threads. Panels (a1)--(a4) display a time sequence of images for Jet3, the white box in panel (a1) outlines the field of view corresponding to panels (b1)--(b2). The green line labeled S1 in panel (a2) traces the trajectory of the filament internal flows and is used to generate the time-distance diagram in panel (c). In each panel, blue curves represent westward flows driven by Jet3, while red curves denote eastward flows. The yellow curve in panel (a3) marks the western footpoints of the filament. Panels (b1)--(b2) show AIA 193 \AA~images, where the dashed line and dotted lines illustrate the movement of a filament thread, and the red horizontal line labeled with S2 is used to make the time-distance diagram in panel (d). Panel (c) presents the time-distance diagram derived from curve S1, with white dashed lines measuring the speeds of westward and eastward flows. Panel (d) displays the time-distance diagram obtained from the horizontal blue line S2 in panel (b2), which denotes the inflation path of the magnetic flux rope associated with Jet3, which expands radially at a rate of \speed{6.5}. Additionally, red sinusoidal curves in panel (d) correspond to the trajectories of rotational motions of the filament threads. Panel (e) provides a schematic of two scenarios depicting different types of filament rotational motions, encompassing mass flows along filament field lines and the untwisting of the filament. The labels P and N in panel (e) indicate the positive end and the negative end of a dextral filament, and the green curved arrow marks the rotational direction of the filament mass.}
\label{fig3}
\end{figure*}

The first row of \nfig{fig3} shows the filament activities driven by Jet3 in the AIA 193 \AA~images (see also the online animation). Jet3 occurred at 00:28 UT at the same location as Jet1 and Jet2, but it ejected more violently than the two previous jets and developed well at 00:56 UT. The ejecting hot plasma entered the filament flux rope at a higher temperature, illuminating the whole filament flux rope structure. The hot jet plasma moved westward along the filament flux rope as indicated by the blue arrow in \nfig{fig3} (a2) -- (a4), which caused an arc-shaped brightening region at the west footpoint of the filament due to the pile-up of hot plasma (see the yellow curve in \nfig{fig3} (a3)). After this, we observed some eastward mass flow along the filament as indicated by the red arrow in \nfig{fig3} (a4). To study the detailed kinematics of the internal mass flows along the filament, we made a time-distance diagram along the filament axis (see S1 in \nfig{fig3} (a2)), and the result is plotted in \nfig{fig3} (c).  It is measured that the westward mass flow was at a speed of about \speed{122}, while that for the eastward flow was about \speed{110}. Similar to Jet1 and Jet2, the filament threads also exhibited a lateral expansion during the passage of Jet3 (see the red arrow in \nfig{fig3} (b1)). \nfig{fig3} (d) is the time-distance diagram made along a path across the filament (see the red line in \nfig{fig3} (b2)), which can be used to diagnose the lateral expansion and unwinding motion of the filament. Similarly, the filament flux rope also showed an obvious lateral expansion at a speed of about \speed{6.5} (see yellow dashed lines in \nfig{fig3} (d)). In addition, the filament threads showed a sinusoidal shape motion in the time-distance diagram (see the red dotted lines in \nfig{fig3} (d)). According to \cite{2016ApJ...831..126O}, the apparent rotation motion of mass flows in a filament magnetic flux rope is quite different in the untwisting and simple flowing stages along the magnetic structure. Here, we draw a cartoon to demonstrate the two types of rotational motions in \nfig{fig3} (e), using the present dextral filament as an example. In a low $\beta$ plasma environment, the plasma is well frozen on the magnetic field lines. As the plasma moves along the twisted magnetic threads of the filament, its rotational trajectory follows the direction of the filament's twist, i.e., in the counterclockwise direction when looking from the positive end (as indicated by the {green} arrow in the left panel of \nfig{fig3} {e}). However, for an untwisting dextral filament magnetic flux rope, the plasma motion should be in the opposite direction to the twist along with the unwinding filament threads, i.e., in the clockwise direction (as indicated by the {green} arrow in the right panel of \nfig{fig3} (e)). Because the rotational directions of the filament threads in Jet1 to Jet3 were all opposite to the twist direction of the dextral filament, the rotational motion of the filament flux rope should represent the untwisting motion of the filament instead of simple mass flows along the filament magnetic structure. In addition, the appearance of paired blue- and red-shift features during Jet1 and Jet2 can also be explained by the untwisting motion of the filament magnetic flux rope \citep[e.g.,][]{2023ApJ...942L..22D,2021ApJ...911...33C}. Mass flows within filaments exhibit a spiral trajectory with a constant radius. However, in the present study, the filament magnetic flux rope showed an obvious expansion because of the gradual increase of the filament radius \citep{2016ApJ...831..126O}. Here, the expansion and untwisting of the filament might be caused by the reconfiguration of the filament magnetic structure owning to the magnetic reconnection which produced the observed jets near the east footpoint of the filament since the magnetic reconnection between some external magnetic system and the filament magnetic field can not only result in the reconfiguration of the filament magnetic structure but also the redistribution of magnetic twists in the two magnetic systems \citep[e.g.,][]{2011ApJ...735L..43S, 2022MNRAS.516L..12T, 2023MNRAS.520.3080T}. The observed backward mass flow in the filament should be along the laterally expanded portion of the magnetic flux rope because it often appeared after the forward-moving jet plasma flow.

\begin{figure*}
\centering
\includegraphics[width=0.9\textwidth]{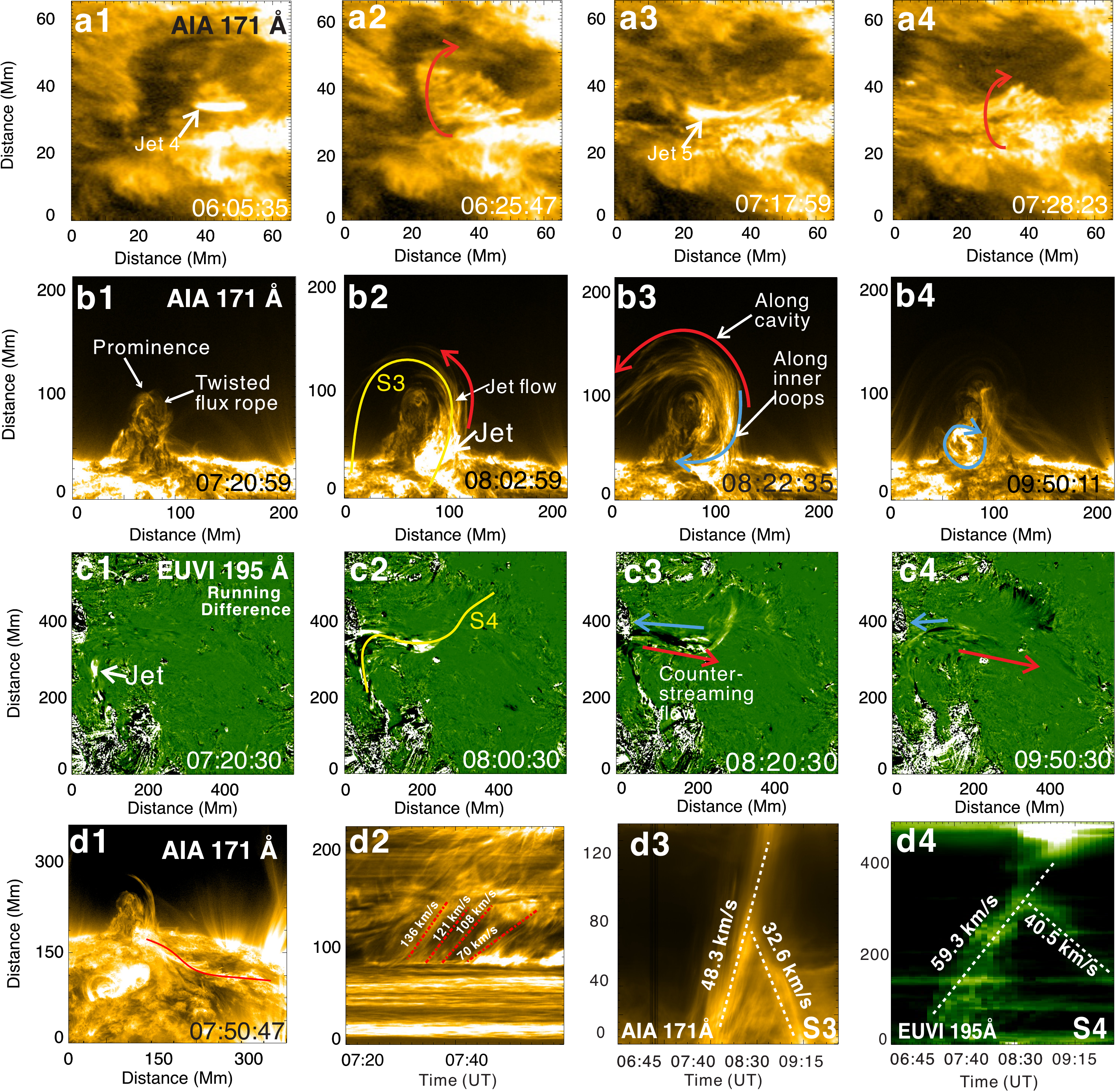}
\caption{The detailed kinematic analysis of the notable jets at the western limb of the solar disk on August 4, 2012. Panels (a1)--(a4) show the beginning and developing stage of the Jet4 and Jet5 in AIA 171 \AA\ images; the red curved arrows denote the rotational direction of the filament threads, which shows clockwise rotation. Panels (b1)--(b4) are AIA 171 \AA~time sequence images for the prominence flows during Jet5, the yellow curve labeled with S3 in panel (b2) was used to make the time-distance diagram in panel (d3), in the second-row panel, the blue curved arrow represents prominence upward flows driven by Jet5. In contrast, the red curved arrow denotes the prominence of downward flows. Panels (c1)--(c4) show the STEREO-A 195 \AA~running difference images for the solar disk filament during the Jet5, the yellow curve labeled with S4 in panel (c2) was used to make the time-distance diagram in panel (d4), In the third-row panel, the blue arrow represents the simultaneous filament westward flows driven by Jet5. In contrast, the red curved arrow denotes the simultaneous filament eastward flows. The red curve in panel (d1) shows the kinematic path of Jet5, and panel (d2) shows the time–distance diagram of AIA 171 \AA\ images along the red curve in panel (d1). Panels (d3)--(d4) show time–distance diagrams made by the slices S3 and S4 (An animation of this event is available).}
\label{fig4}
\end{figure*}

Due to the rotation of the Sun, the filament under study was observed as a prominence at the west limb of the solar disk on August 4, 2012, in the field of view of the {\em SDO}. In addition, since the separation angle between the {\em STEREO}-A and the {\em SDO} was approximately 121.9 degrees on that day, the prominence can be observed as a filament on the solar disk when observing from the {\em STEREO}-A. Although it had been five days, one can also observe that many jets occurred at the eastern footpoint of the prominence, and similar jet-driven mass flows in the prominence. This suggests that the magnetic structure and the magnetic condition of the filament did not change significantly compared to those when the filament was on the disk several days before. We select two conspicuous jets for a detailed analysis (hereafter Jet4 and Jet5), and the results are displayed in \nfig{fig4}. For better visualization, all AIA 171~\AA\ images in the \nfig{fig4} are rotated 90 degrees counterclockwise. Jet4 (see \nfig{fig4} (a1)) and Jet5 (see \nfig{fig4} (a3)) occurred at about 06:05 UT and  07:17 UT, respectively. During the periods of the two jets, the clockwise rotation motion of the prominence threads can be observed (see the red arrows in \nfig{fig4} (a2) and (a4)), indicating the untwisting motion of the prominence (see the videos available in the online journal). We further examined the evolution of the associated mass flows launched by Jet5 in the second and third rows of \nfig{fig4}. At the beginning of Jet5 at about 07:20 UT, meanwhile, the prominence exhibited as a twisted magnetic flux rope (see \nfig{fig4} (b1) and (c1)),  we measured the speed of Jet5 along its trajectory (see the red line in \nfig{fig4} (d1)) by the time-distance diagram, the measured speeds range from \speed{70} to \speed{136} (see red dashed-dotted lines in \nfig{fig4} (d2)). When the jet plasma reached the middle section of the filament observed by the {\em STEREO}-A, it manifested a dominant forward mass flow along a circular prominence cavity structure in the corresponding AIA 171 \AA\ image (see \nfig{fig4} (b2)). At about 08:23 UT, due to the appearance of a backward mass flow parallel to the forward one, it formed a picture of counterstreaming mass flow in the filament on the {\em STEREO}-A images (see \nfig{fig4} (c3)). In the corresponding AIA 171 \AA\ images, it can be identified that the backward mass flow was the downward moving plasma flow formed at a projection height of about 60 Mm above the solar surface near the apex of the prominence flux rope (see the blue arrow in \nfig{fig4} (b3)). In addition, it can be distinguished that the forward and downward mass flows were along the outer and inner circular magnetic threads of the prominence magnetic flux rope. Two time-distance diagrams are created along the yellow curves shown in \nfig{fig4} (b2) and (c2), and the results are plotted in \nfig{fig4} (d3) and (d4), respectively. The projection flow speeds measured from the time-distance diagrams suggest that the upward (downward) mass flow in the AIA 171 \AA\ images was about \speed{48.3 (32.6)}. In contrast, the corresponding forward (backward) mass flows in the {\em STEREO}-A images were about \speed{59.3 (40.5)}.

\begin{figure*}
\centering
\includegraphics[width=0.9\textwidth]{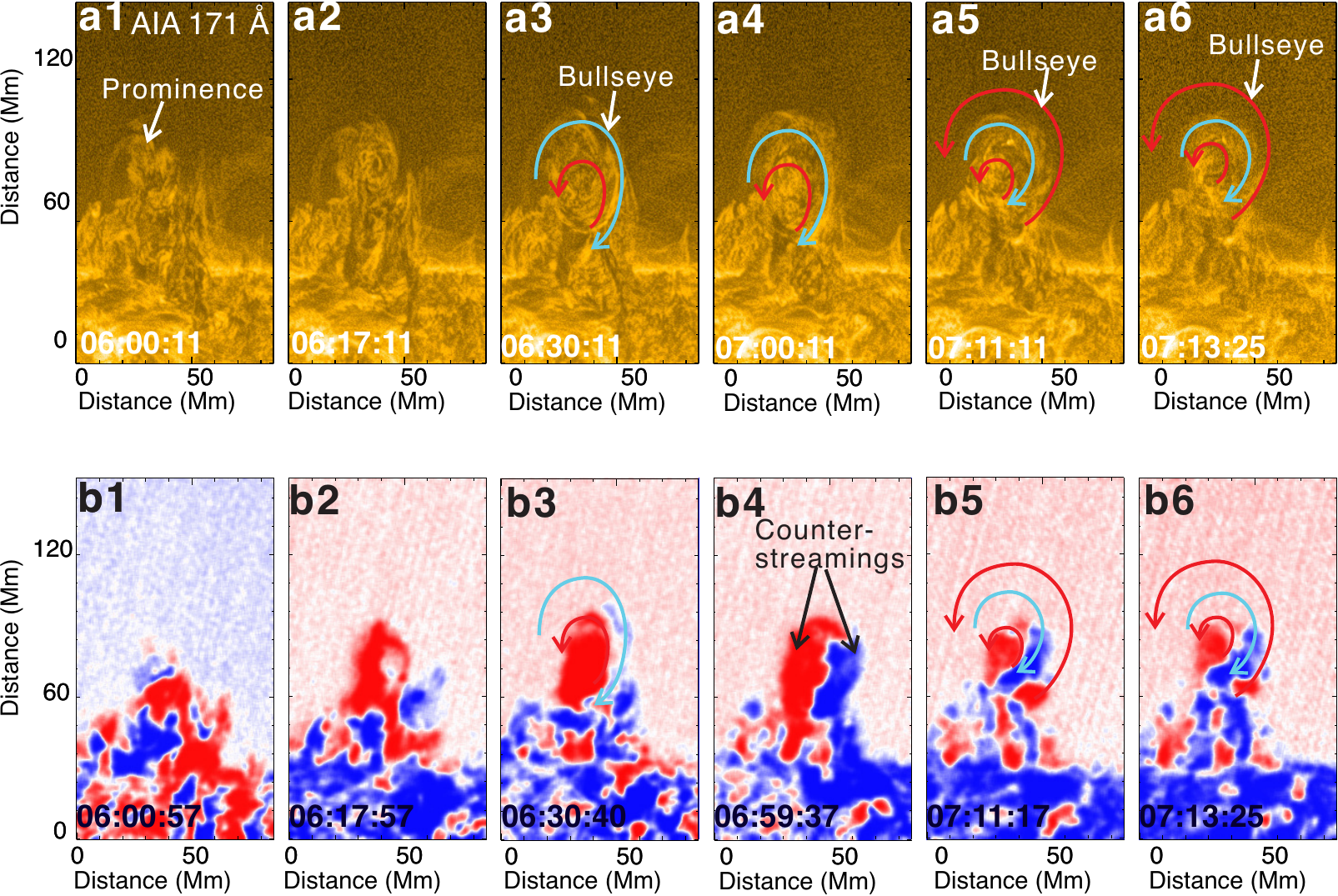}
\caption{The formation of prominence counterstreaming mass flows during the Jet4. Panels (a1)--(a6) show the AIA 171~\AA~images for the prominence during the Jet4. Panels (b1)--(b6) show the time sequence images of the pseudo-Doppler images, which are created by subtracting H$\alpha$-0.5~\AA~images from the paired near-simultaneous H$\alpha$+0.5~\AA~ones. In the panels, the blue curved arrow represents the prominence's downward mass flows driven by Jet5, while the red curved arrow denotes the prominence's upward mass flows.}
\label{fig5}
\end{figure*}

The circular counterstreaming mass flows in the prominence remind us of the particular Doppler bullseye pattern associated with coronal cavities \citep{2013ApJ...770L..28B}. Unfortunately, no Doppler velocity observations were available on 2012 August 4, such as the CoMP. Therefore, we use AIA 171~\AA~images and alternatively generate pseudo-Doppler velocity images using the near-simultaneous H$\alpha$ line-center and off-band observations the SMART takes. The top row of \nfig{fig5} shows the  AIA 171 \AA\ {images}. As indicated by the circular blue and red arrows, concentric {alternate pattern of} blue- and red-shift Doppler velocity can well be identified, and they together form a concentric bullseye pattern resembling those observed by the CoMP \citep{2013ApJ...770L..28B}. Here, the composite images also indicate that the bullseye pattern corresponds to the prominence magnetic flux rope cross-section. The pseudo-Doppler images displayed in the second row of \nfig{fig5} are created by subtracting H$\alpha$-0.5 \AA\ images from the paired near-simultaneous H$\alpha$+0.5 \AA\ ones. The Doppler velocity images revealed a pair of blue- and red-shift velocities in the line of sight, suggesting simultaneous counterstreaming mass flows in the prominence moving toward and away from the observer. We note that the locations of the blue- and red-shift mass flows were consistent with the circular counterstreaming mass flows identified in the tri-color composite images (see the circular arrows in \nfig{fig5} (b5) and (b6)).

\begin{figure*}
\centering
\includegraphics[width=0.9\textwidth]{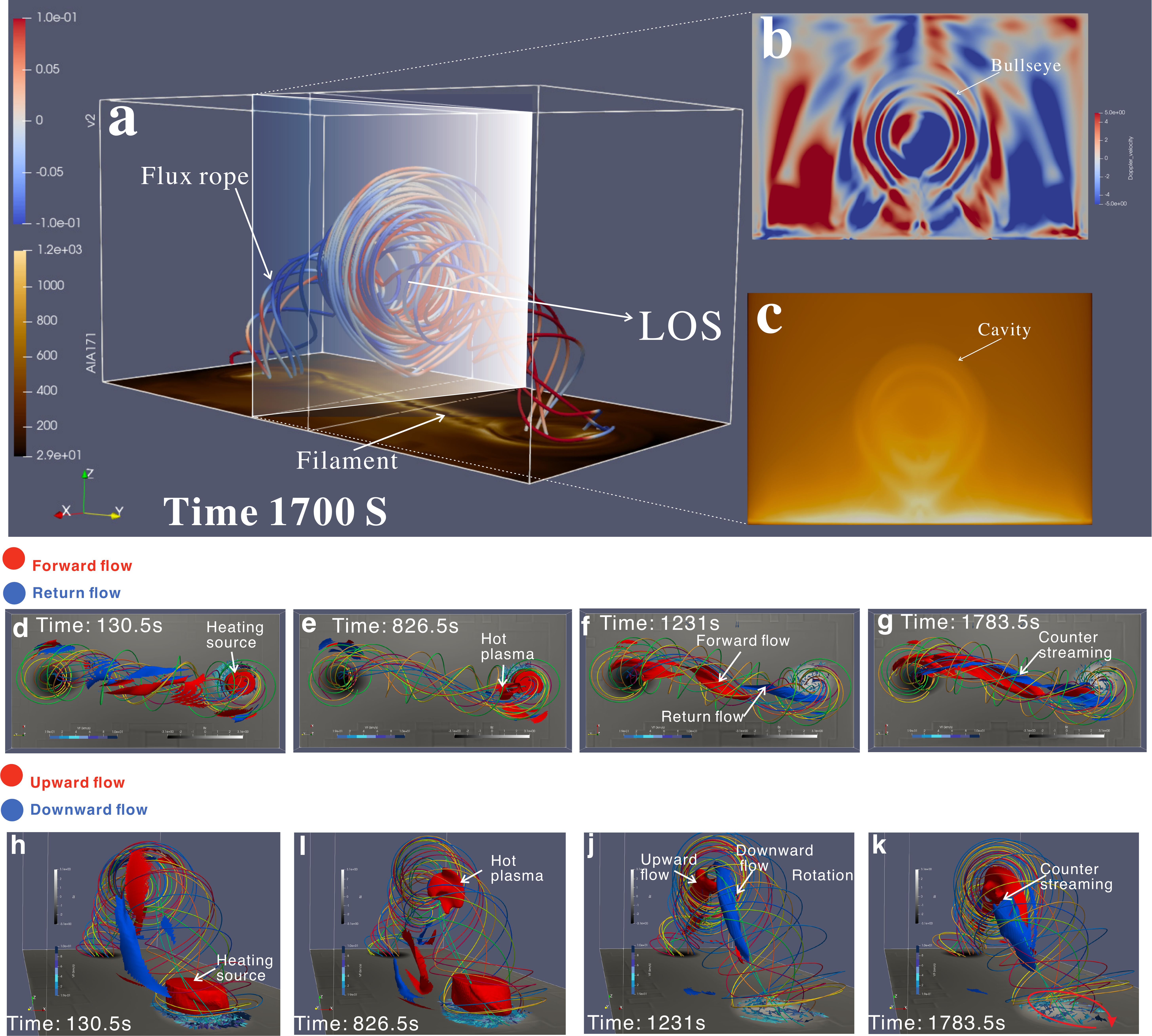}
\centering
\caption{The detailed information for the filament counterstreaming mass flows and the prominence bidirectional mass flows in our simulation work. Panel (a) displays the magnetic flux rope of the filament at 1700 seconds. Panel (b) shows the simultaneous synthesis of the Doppler velocity map, which exhibits a clear bullseye pattern. Panel (c) shows the simultaneous AIA 171 \AA\ synthesis image; the cavity can be seen in the synthesis image. Panels (d)--(g) show the formation of filament counterstreaming mass flows observed on the solar disk, where the arrows mark the flow patterns (forward and return flows) in the filament view.  Panels (h)--(k) show the formation of prominence bidirectional flows observed at the solar limb, where the arrows indicate the flow patterns (upward and downward flows) in the prominence view. In the panels, the motions of filament footpoints are indicated by blue arrows, with darker shades indicating higher velocity. The red curved arrow in panel (k) shows the trajectory of the filament footpoints.
\\(An animation of this simulation is available)}
\label{fig6}
\end{figure*}

\section{Numerical modeling}\label{sec:numerical}
We perform a three-dimensional numerical simulation using the AMRVAC code to verify our observations \citep{2018ApJS..234...30X, 2023A&A...673A..66K}. In the simulation, a regularized Biot-Savart Law (RBSL) magnetic flux tube is assumed as the filament magnetic flux rope \citep{2021ApJS..255....9T}, and the atmospheric region comprises a chromosphere of thickness 500 Km and a corona extending up to 200 Mm. Below are the detailed settings in our simulation.

Previous research has examined the fine structure of filaments using various models. Here, we follow the methodology in \citet{2014ApJ...790..163T}, where the bottom boundary temperature decreases exponentially to achieve hydrostatic equilibrium within a 1 MK isothermal corona. The temperature distribution of the isothermal atmosphere, including the idealized chromosphere and corona, is based on the approach outlined by ~\citet{2016ApJ...823...22X}. A localized heating at the footpoint of the magnetic flux rope system is used to produce coronal jets along the filament magnetic flux rope.

In our simulation, the integral axis path of the flux ropes follows a circular arc shape in the vertical plane of the configuration. The great-circle distance of the arc is 113.731 Mm, the shortest-circle distance is 30 Mm, and the torus minor radius is set to 30 Mm. Consistent with \cite{2016ApJ...823...22X}, the density of the filament is concentrated in the concave-up dips of the flux rope structure. We directly attach the filament density to these concave-up dips within the flux rope structure. The functions are below.

\begin{equation}
dip=\nabla B_z\cdot\mathbf{\hat{B}}
\end{equation}

\begin{equation}
\begin{split}
\rho_{prom}=2.341668\cdot10^{-15}\cdot\left(12-0.3\cdot z\right) g/cm^{3}\\
 \left(3 Mm \le z \le 15 Mm, dip \ge 0, \frac{B_z}{B} \le 0.008\right) \\
\end{split}
\end{equation}

The initial background heating is given by

\begin{equation}
\bm{\hat{b}} = \frac{\bm{\hat{B}}}{\|\bm{\hat{B}}\|}
\label{eq:unit_b}
\end{equation}

\begin{equation}
\bm{\kappa} = \bm{\hat{b}} \cdot \nabla \bm{\hat{b}}
\label{eq:kappa_def}
\end{equation}

\begin{equation}
\alpha = 
\begin{cases}
0.1, & \text{if } \|\kappa\| < 0.1 \\
1.0, & \text{if } \|\kappa\| > 1.0 \\
\|\kappa\|, & \text{otherwise}
\end{cases}
\label{eq:kappa_limit}
\end{equation}

\begin{equation}
	\mathcal{H}_0 = 10^{-4}\left(0.5\times B^{1.56} \alpha^{0.75}\rho^{0.125}+0.5\times\exp\left( -\frac{z}{20} \right) \right)\text{erg cm}^{-3}.
	\label{eq:bh}
\end{equation}

To model the global coronal heating responsible for maintaining a hot corona, we implement a mixed heating model~\citep{2022A&A...668A..47B} (see Equation~\ref{eq:bh}). One of the terms of the equation follows the prescriptions of~\citet{2013ApJ...773..134L} and~\citet{2016ApJ...817...15M}, where $\kappa$ is the local curvature of the magnetic field. Considering the energy transport from lower to upper atmospheric layers, it is hypothesized that the heating efficiency attenuates with increasing altitude~\citep{1981ApJ...243..288S,2002ApJS..142..269A}. Another term is defined as a multivariate power-law.

 {In our numerical model, the chromosphere is assigned a thickness of $500\,\text{km}$. Following the previous works, we set localized heating in a Gaussian function of distance from the peak position, $s_{peak}$~\citep{1999ApJ...512..985A, 2022A&A...663A..31N}. To restrict the localized heating at the low corona, we set $s_{\text{peak}} = 2.1\,\text{Mm}$. Additionally, $t_{\text{peak}} = 1200\,\text{s}$ is adopted to control
The periodicity of the localized heating, which directly corresponds to the jet lifetime of 1200 seconds \citep{2014A&A...567A..11Z,shen21}.}

\begin{flalign}
    & H_l(s,t) = 10^{-3} \exp\left[ -(s - s_{\text{peak}})^2 \right] \left| \sin\left( \frac{\pi t}{t_{\text{peak}}} \right) \right| \, \mathrm{erg\,cm^{-3}}, & \nonumber \\
    & \quad t < t_{\text{peak}},\ B_z \geq 6\,\mathrm{G} & \label{eq5}
\end{flalign}

The numerical simulation results are displayed in the \nfig{fig6}, which well reproduced the dextral filament magnetic flux rope (\nfig{fig6} (a)), the Doppler bullseye pattern in the synthesis Dopplergram with three {alternate pattern of} blue- and red-shift circular rings (\nfig{fig6} (b)), and the tunnel structure observed in synthesis EUV 171 \AA~images (\nfig{fig6} (c)). The bullseye pattern in the synthesis Dopplergram has a velocity of about ten \speed{}, which is of a similar order of magnitude as the speed measured from CoMP observations.

As the jet-driven counterstreaming mass flows are observed to be very different in filament and prominence, we select two different observing angles to show the generation and evolution of the counterstreaming mass flows in the filament and prominence, as respectively displayed in the middle and the bottom rows of \nfig{fig6}. The simulation successfully reproduces the counterstreaming mass flow pattern in the prominence; we focus on a pair of counterstreaming mass flows in the simulated magnetic flux tube, where the upward ($V_{y} ~\textgreater ~$\speed{20}$~ $ and $T  ~\textgreater ~1.4 ~{\rm Mk}$) and downward ($V_{y}  ~\textless~ $\speed{-20}$ $ and $T~ \textgreater~ 1.4~ {\rm Mk}$) plasma flows are indicated by the red and blue colors in \nfig{fig6} (d) -- (k). 

The middle row shows the generation of the counterstreaming mass flow in a filament, i.e., the top view of the magnetic flux rope. {The setting of localized heating at the footpoint of the flux rope in our simulation work drives the ascent of the local plasma into the flux rope, which can be regarded to be an effective alternative way to produce the upward flows in the observational jet events (see the labels hot plasma in \nfig{fig6}). It can be observed that due to the heating at the right footpoint of the filament, the heated hot plasma flow starts to move along the filament spine toward the left (\nfig{fig6} (d) and (e)). As the leftward mass flow approaches the middle part of the filament, a rightward-moving plasma flow is generated (blue, \nfig{fig6} (f)). Since these two mass flows move in opposite directions, they together form the expected counterstreaming mass flow pattern in the filament, as revealed in our observation (\nfig{fig6} (g)). For simulating the counterstreaming mass flows observed in prominences, we display the axis view of the magnetic flux rope and mass flows in the bottom row \nfig{fig6} (f), and the images plotted in panels \nfig{fig6} (h) -- (k) are at the same time corresponding to those plotted in \nfig{fig6} (d) -- (g). It can be observed that due to the heating at one of the footpoints of the prominence, hot plasma flow starts to rise and forms an upward mass flow (red) in the prominence (\nfig{fig6} (h) and (i)). As the upward mass flow gets close to the apex of the prominence, a downward mass flow (blue) forms and starts to fall back. This naturally forms the pattern of counterstreaming mass flows or the Doppler bullseye pattern observed in our observation (\nfig{fig6} (k)).
   
\section{Conclusions and Discussions}\label{sec:summary}
Using stereoscopic high spatiotemporal resolution observations provided by space and ground-based solar telescopes, including the {\em SDO}, the {\em STEREO}-A, the SMART, and the CoMP, we firstly investigated the origin and evolution of counterstreaming mass flows simultaneously observed from two different view angles based on a long-lived transequatorial filament from July 23 to August 4, 2012. In addition, we also performed a three-dimensional MHD simulation to test our observational results. Both our observational and numerical simulation results show that counterstreaming mass flows in on-disk filaments and limb prominences can be launched by repeated coronal jets occurring at one of the footpoints of filaments or prominences. {Our results align with prior studies. Firstly, our counterstreamings are initiated by eruptive events at the filament footpoint, with coronal jets acting as the driving force to inject mass and momentum into filaments or prominences, leading to flows~\citep{2019ApJ...881L..25Y,2020ApJ...897L...2P}. Secondly, the flow characteristics, such as the velocity range and the temporal evolution pattern, show similarities. Our counterstreamings move within the range of about \speed{40--110}, similar to that seen in the EUV observations~\citep{2013ApJ...775L..32A,2022A&A...659A.107C}. Finally, our study's spatial distribution of the counterstreaming flows shares resemblances with previous findings. The flows are typically concentrated along the axis of the filaments or prominences, with a clear separation of the two moving directions, similar to the patterns observed by~\citet{2019ApJ...881L..25Y} and \citet{1998Natur.396..440Z}. Our study also indicated that the recently reported Doppler bullseye pattern associated with coronal cavities is the manifestation of the counterstreaming mass flows in prominences, which indirectly exhibit the helical magnetic structure of prominence magnetic flux ropes. In addition, during the generation and evolution of the counterstreaming mass flows, obvious lateral expansion and rotation motion of the filament is observed, which might reflect the reconfiguration of the filament magnetic field caused by the intruding of coronal jets and redistribution of magnetic twists between the filament and the external magnetic system that produces the observed coronal jets along the filament through magnetic reconnection between them. {Additionally, our counterstreaming model reveals 3D dynamical effects in the filament's internal flows, including lateral expansion and rotational motion within the flux rope, which also enhances the understanding of the bullseye phenomenon in prominence.}

Counterstreaming mass flows are ubiquitous in every filament and prominence \citep{1998Natur.396..440Z, 2003SoPh..216..109L}, but the question of their origin is still hotly debated. Previous studies proposed several candidate mechanisms for interpreting the generation of counterstreaming mass flows in filaments, which have been briefly reviewed in Section \ref{intro}. Here, we propose a new physical mechanism for accounting for the generation of large-scale counterstreaming mass flows in filaments based on our observational and numerical simulation results. We argue that counterstreaming mass flows can be generated by repeated coronal jets at one of the footpoints of a filament consisting of helical magnetic field lines (i.e., a magnetic flux rope). It starts with the occurrence of small-scale magnetic activities (e.g., flux emergence) close to one of the footpoints of a filament magnetic flux rope, which reconnects with the filament magnetic field and results in a hot coronal jet traveling along the helical magnetic field lines of the filament, which can be observed as a forward moving hot plasma flow. In the meantime, the reconnection causes the redistribution of magnetic twists between the filament and the small-scale magnetic field system and the reconfiguration of the filament's magnetic structure, causing observable characteristics such as lateral expansion and untwisting motions of the filament. As the continuous traveling of the jet plasma, it will go through the concave bottom and convex apex of the filament magnetic flux rope periodically. As a magnetic flux rope is characterized by a higher plump body but lower thin ends rooted in the chromosphere, during the rising phase of a moving plasma from one end of a filament, it needs more and more kinetic energy to overcome the higher and higher gravitational potential energy to pass over the next apexes of the helical magnetic field line. If a moving plasma can pass through the highest apex of the filament, it would undergo a process that is exactly opposite to the rising phase. In addition, the ejecting hot jet plasma will gradually lose its kinetic energy for some reason, such as plasma viscosity. This can also lead to the failed passage of the ejecting plasma over some apex of the helical magnetic field of the filament due to insufficient kinetic energy, even if its initial kinetic energy is enough. Because of these reasons, some moving plasma will fall back at a location close to some apex to become a backward-moving plasma flow in the filament. The simultaneous forward and backward mass flows naturally form the picture of counterstreaming flows in filaments and the Doppler bullseye pattern observed in prominences. 

By ignoring the effects of pressure gradient force, the magnetic forces, and the viscous friction, we can roughly estimate how fast a jet should be to pass through an apex of the helical magnetic field lines in a filament by simply considering the conservation of the jet's kinetic energy and gravitational potential energy, i.e., $mg{h}=\frac{1}{2}mv^2$. Here, $g$ is the acceleration of the solar gravity, $m$, $h$, and $v$ are the jet plasma's mass, height, and initial velocity, respectively. Taking Jet5 as an example, the projection height of the prominence's apex was at least 110 Mm. Therefore, if the ejecting jet plasma passes the apex of the prominence, it needs an initial velocity of at least about $v = \sqrt{2gh}$ = \speed{245}. The measured projection speed of Jet5 based on the AIA 171 \AA\ images is about \speed{136}. If we consider the inclination angle of the filament is within an angle of 30 -- 60 degrees concerning the solar surface, the true velocity of the jet should be within \speed{157 -- 272}. The least required velocity of about \speed{245} is within the estimated velocity range of Jet5, consistent with our observational results that some of the ejecting jet plasma passed over the prominence apex and continued to move forward. In the meantime, some of the slower ejecting plasma of the jet started to fall back at a height of about 60 Mm above the solar surface. It should be pointed out that Jet5 is a stronger jet during our observation time, and the velocities of the vast majority of other jets are slower than Jet5. Therefore, a large proportion of the observed jets have transformed into backward mass flows before they reached the apex at a height of 110 Mm, even though they passed other relatively lower apexes in the filament's helical magnetic field lines.

The counterstreaming flows we found appear to be distinct from those formed by the longitudinal oscillation of filament plasma within magnetic dips \citep[e.g.,]{1997SoPh..172..181M, 2003SoPh..216..109L, 2006SoPh..236...97J}. In these studies, the countersteaming flows are formed by longitudinal oscillations out of phase along different magnetic field lines. In contrast, our study's mass flows were moved unidirectionally and did not oscillate within filament dips. \cite{2014ApJ...784...50C} suggested that H$\alpha$ counterstreaming mass flows are caused by the combination of longitudinal oscillations in magnetic dips and alternating unidirectional flows along different filament threads. Their theoretical analysis suggested that the siphon mechanism can form the mass flow due to the imbalance of magnetic field strength and mass density at the two footpoints of a filament thread, and counterstreaming mass flows in a filament can be formed when mass flows in opposite directions in different filament threads. Our mechanism differs from \cite{2014ApJ...784...50C}. Firstly, the origin of the mass flow in our mechanism is due to the injection of hot jet plasma at one footpoint of the filament, rather than due to siphons. Secondly, the return mass flow formation is due to the partial fallback of the forward-moving ejecting plasmas due to insufficient kinetic energy rather than unidirectional mass flow along another filament thread. We note that \cite{2010ApJ...721...74A} reported the appearance of return flow in prominence when some of the moving plasma fragments approached the endpoint of the prominence spine close to the associated active region. The return flow is different from the backward mass flow found in our event in several aspects. Firstly, it occurs close to the endpoint of the prominence rather than the apex of the helical magnetic field lines; therefore, the restoring force of the return flow might be magnetic, arising from either magnetic tension or magnetic pressure gradient rather than gravity. Secondly, for prominence with a loop-like solenoid magnetic structure, there are multiple apexes distributed at different heights, and mass flows with different kinetic energy can fall back to become backward flows at different heights. Whereas the return flow found in  \cite{2010ApJ...721...74A} can only occur close to the endpoint of the prominence owing to the magnetic mirror effect. Thirdly, the speeds of the counterstreaming mass flows in our event (\speed{40--110}) are faster than those (\speed{10--20}) in \cite{2010ApJ...721...74A}. This suggests their different origins. The counterstreaming mass flows in our event were caused by coronal jets at one footpoint of the prominence, while those in \cite{2010ApJ...721...74A} might have resulted from the net force resulting from a small deviation from magnetohydrostatic equilibrium.

During the traveling of the ejecting jet plasma within the filament magnetic flux rope, the filament exhibited obvious lateral expansion at a speed of a few \speed{}. This might be due to the transient increased pressure inside the rope because every coronal jet ejected a certain amount of mass into the filament. As a natural response to the disturbance, the filament magnetic flux rope would decrease its mass density to keep a force balance through expansion to increase the volume. However, the expanded magnetic flux rope should be unsustainable since the increased mass will drain back to the solar surface along its two legs. Although we do not observe in the present study, the filament magnetic flux rope will inevitably experience a shrinking phase after the expansion.

High-resolution observations reveal that rotational motions play an important role in the dynamic of the filament's fine structure; they are directly relevant to the accumulation and release of magnetic energy and play a key role in understanding the evolution of magnetic flux ropes ~\citep{2016ApJ...831..126O, 2024SCPMA..6759611Z,2024ApJ...974L...3Z,2023ApJ...953L..13L}. As discussed in Section \ref{sec:result}, we argue that the observed rotation of the filament flux rope might be caused by the magnetic reconnection that produced the repeated coronal jets. Since the reconnection occurred between the filament and the small-scale magnetic system below it, it can naturally result in the untwisting and reconfiguring of the filament's magnetic field, resembling the physical mechanism used to explain unwinding coronal jets in open magnetic field lines \citep[e.g.,][]{2011ApJ...735L..43S, 2019ApJ...873...22S}. Other possibilities might also cause the untwisting phenomenon of the filament. For example, the magnetic reconnection between the filament and the background confining magnetic field. However, we do not find relevant observational evidence to support such a scenario. In addition, the lateral expansion can also lead to the untwisting phenomenon of the filament magnetic flux rope.

Filaments/prominences are often interpreted as magnetic flux ropes—twisted structures of magnetic field lines that could potentially store and release significant amounts of energy~\citep{rust96, zhangj12, Zhou2017, 2022ApJ...934L...9L, 2024ApJ...964..125S,2024ApJ...965..160C,2023ApJ...952...43Y}. However, the magnetic flux rope structure of filaments/prominences remains an ambiguous and intriguing aspect of solar physics. Despite numerous observational and simulation works have widely used the conclusion that filaments/prominence are magnetic flux ropes, direct observational evidence that unequivocally links the filament or prominence to flux ropes is scarce, with only some indirect evidence \citep{2013ApJ...770L..28B,2016ApJ...831..126O,2017ApJ...851...67S,2019ApJ...883..104S}. The present study presents a good observational example that can support the filaments' magnetic flux rope scenario, with multiple pieces of evidence, including the fine helical structure traced out by the coronal jets, the untwisting motion of the filament, and the Doppler bullseye pattern. In addition, our observation also provides a clear paradigm of the expected physical picture that the mass of filaments can be obtained by the direct injection of coronal jets \citep{1999ApJ...520L..71W, 2003ApJ...584.1084C, 2019ApJ...883..104S, 2021ApJ...913L...8H}.

\begin{acknowledgments}
The authors are grateful for the excellent data the SDO, STEREO, SMART, and CoMP teams provided. The authors gratefully acknowledge the reviewer's constructive comments and insightful suggestions, which have significantly improved the quality of this work. This work is supported by the Natural Science Foundation of China (12173083), the Shenzhen Key Laboratory Launching Project (No. ZDSYS20210702140800001), the Strategic Priority Research Program of the Chinese Academy of Sciences (XDB0560000), and the Key Research Program of Frontier Sciences, CAS (grant No. ZDBS-LY-SLH013). The numerical computations were conducted on the Yunnan University Astronomy Supercomputer.

\end{acknowledgments}

\end{document}